\pdfoutput=1
\documentclass{iau} 
\usepackage{graphicx}
\usepackage{natbib}
\usepackage[suffix=]{epstopdf}
\usepackage{aas_macros}

\title[IAUS 320.~~White-Light Continuum in Stellar Flares] 
{White-Light Continuum in Stellar Flares}

\author[Adam F. Kowalski]   
{Adam F. Kowalski$^{1,2}$}

\affiliation{$^1$Department of Astronomy, University of Maryland,
  Stadium Drive, College Park, MD 20742, USA \\ email: {\tt adam.f.kowalski@nasa.gov} \\
  $^2$NASA Goddard Space Flight Center, Heliophysics Science
Division, Code 671, 8800 Greenbelt Rd., Greenbelt, MD
20771, USA.
}

\pubyear{2015}
\volume{320}  
\jname{Solar and Stellar Flares and Their Effects on Planets}
\editors{A.G. Kosovichev, S.L. Hawley, P. Heinzel}
\begin{document}

\maketitle

\begin{abstract}
In this talk, we discuss the formation of the near-ultraviolet and optical continuum emission in M dwarf flares through
 the formation of a dense, heated chromospheric condensation.   Results are used from a recent radiative-hydrodynamic model of the response of 
an M dwarf atmosphere to a high energy flux of nonthermal electrons.   These models are used to infer
the charge density and optical depth in continuum emitting flare layers from spectra covering the Balmer jump and optical wavelength regimes.  
Future modeling and observational directions are discussed. 

\end{abstract}

\firstsection 
\section{Introduction}
M dwarf flares produce bright white-light (broadband)
continuum emission, which is most easily characterized at the near-ultraviolet (NUV) and
blue optical wavelengths because of the large luminosities
and the high contrast against the background stellar photosphere.
The impulsive phase NUV and optical broadband color distribution exhibits the trend of a hot
($T\approx9000-14,000$ K) blackbody spectrum \citep{HawleyPettersen1991, HawleyFisher1992, Hawley2003, Zhilyaev2007, Lovkaya2013}, which is not readily produced
by any plausible flare heating mechanism thus far considered \citep{HawleyFisher1992}.  Broad wavelength
coverage spectra around the Balmer jump have shown that the impulsive
phase color temperature is consistent with a hot blackbody at
wavelengths $\lesssim3646$ \AA\ and $\gtrsim4000$ \AA, but there is a relatively small
jump in flux in the wavelength range from $\lambda=3646 - 4000$ \AA.  In this spectral region, the higher
order Balmer lines apparently blend together \citep{Zarro1985, Doyle1988, HawleyPettersen1991}, which is a phenomenon that is also 
observed in spectra of solar flares \citep{Donati1985}.

The most widely accepted impulsive heating mechanism in solar flares
is collisional heating by nonthermal electron beams accelerated in the
corona, which was modeled in an M dwarf atmosphere in 
\cite{Allred2006} using beam parameters obtained at the peak of a large solar flare \citep{Holman2003}. 
Specifically, nonthermal electron energy fluxes of $10^{10}-10^{11}$ erg cm$^{-2}$ s$^{-1}$ were modeled with a 
low-energy cutoff of $E=37$ keV in the electron distribution.
 The model NUV and optical continuum spectra were found to exhibit bright chromospheric Balmer continuum emission and a cool color temperature 
at optical wavelengths which is due to chromospheric Paschen continuum emission and enhanced photospheric radiation.
 The continuum flux ratios in these model spectra are 
 inconsistent with the impulsive phase observations of many M dwarf flares \citep{Kowalski2013}, suggesting that alternative
 energy deposition scenarios are required to produce hotter, denser flare atmospheres.
However, a large range of reasonable parameters for nonthermal electron beams has not been investigated with 
radiative-hydrodynamic flare models, and a detailed exploration of the predictions for the extreme values in this range is necessary before
turning to alternative heating mechanisms.  Unprecedented 
energy fluxes of $>10^{12}$ erg cm$^{-2}$ s$^{-1}$ in electron beams have been inferred for the brightest solar flare kernels
from recent high spatial resolution images \citep{Krucker2011, Gritsyk2014, Milligan2014}, and these large
energy fluxes may help explain the flare continuum radiation from stars that are more magnetically active compared to the Sun.

In this paper, we summarize the results of \cite{Kowalski2015} (hereafter K15), 
which presented the atmospheric response to very high nonthermal electron beam fluxes as high as
 $10^{13}$ erg cm$^{-2}$ s$^{-1}$, which is one- to two orders of magnitude higher than 
was previously possible with the available computational resources.  In K15, we found that the instantaneous NUV and optical continuum
distribution after $t\approx2$~s was
consistent with the spectral observations of the impulsive phase of M dwarf flares.  In addition,
we applied a new modeling technique to the spectral region just redward ($\lambda=3646-3730$ \AA) of the Balmer jump  ($\lambda=3646$ \AA)
where the highest order Balmer lines broaden significantly from the Stark effect and form a (pseudo-)continuum.
The hot blackbody-like continuum inferred from broadband color observations was found to originate from a compression of the chromosphere by hydrodynamic shocks and heating
of this region by precipitating nonthermal electrons.  In K15, we
discussed the detailed formation of three wavelengths ($\lambda=$3550 \AA, 4300 \AA, and
6690 \AA) in the F13
model.  In these proceedings, we discuss the formation
of the continuum radiation at all NUV and optical wavelengths in the F13 flare atmosphere.

\section{The Continuum Radiation from the F13 Beam-Heated Atmosphere}
\begin{figure}[b]
\begin{center}
 \includegraphics[width=5.5in]{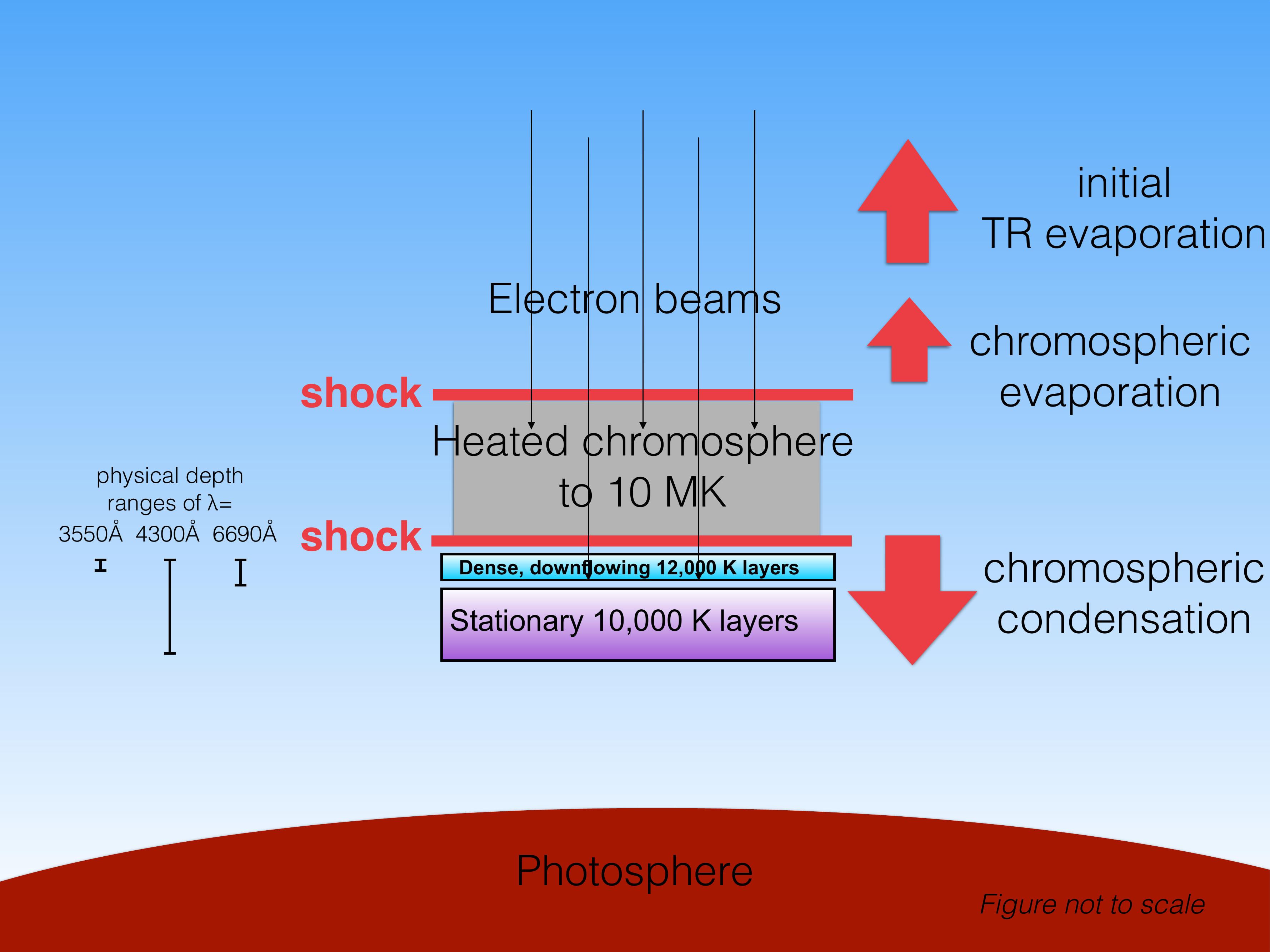} 
 \caption{A qualitative representation of the M dwarf atmosphere in response to the F13 nonthermal electron beam energy deposition.  The 
  physical depth ranges of NUV, blue, and red wavelengths from Figure \ref{fig:panels} (bottom) are indicated.  Note, at $t=2.2$~s, a numerical adjustment
  to the computation results in the upper shock being smoothed over, which allows progression of the simulation.
  For details of the atmospheric evolution, see K15. }
   \label{fig:cartoon}
\end{center}
\end{figure}

In this talk, we consider the atmospheric response of an M dwarf to a nonthermal electron beam with a constant energy flux
of $10^{13}$ erg cm$^{-2}$ s$^{-1}$ (F13) and a double power-law distribution with a minimum (cutoff) energy
of 37 keV.  The simulation was calculated with the RADYN \citep{Carlsson1997} and RH \citep{Uitenbroek2001} codes,
and is described in detail in K15.  
The F13 beam produces two hydrodynamic shocks in
the mid to upper chromosphere, and the thermal pressure from these shocks
drives material upward (chromospheric evaporation) and downward
(chromospheric condensation, hereafter ``CC'').  The flare atmosphere is 
illustrated in Figure \ref{fig:cartoon}.  

The NUV and optical continuum radiation originates from the CC with
densities as high as $n_{e,\rm{max}} \approx 5.6\times10^{15}$ cm$^{-3}$ and from non-moving
(hereafter, the ``stationary'') dense ($n_e\approx10^{15}$ cm$^{-3}$) layers below the
CC.  The CC and stationary layers
are indicated in Figure \ref{fig:cartoon}; the large
number of low-energy electrons in the F13 beam ($E=37-60$ keV) is responsible
for the rapid heating of the upper chromosphere to $T=10$ MK, which occurs within a short
time after helium is completely ionized.  The
higher energy beam electrons ionize and heat the lower atmospheric
heights (the CC and stationary flare layers).

The properties of the surface flux distribution at
$t=2.2$~s are consistent with the impulsive phase constraints
from the spectral flare atlas of \cite{Kowalski2013}.
The F13 surface flux distribution exhibits a Balmer jump
ratio (the ratio of NUV continuum to blue continuum flux)
of $\sim$2 and a color temperature at NUV wavelengths ($\lambda\lesssim3720$ \AA) and
at blue-optical wavelengths ($\lambda=4000-4800$ \AA) of $T\approx10,000$ K, which are typical properties of observed impulsive phase spectra. In contrast, an 
F11 model (also considered in K15) exhibits a Balmer jump ratio of $\sim9$ and a color temperature of $T\approx5000$ K
at blue-optical wavelengths.

To determine the origin of the emergent intensity (and thus the surface flux) at all continuum wavelengths, we use the atmospheric
parameters from RADYN to calculate the contribution
function, $C_I = dI_{\lambda}/dz$, to the emergent intensity:


\begin{equation} \label{eq:didz}
C_I = \frac{dI_{\lambda}}{dz} = \frac{j_{\nu}}{\mu} e^{-\tau_{\nu}/\mu} \frac{c}{\lambda^2}
\end{equation}

which is the total continuum emissivity ($j_{\nu}$) at a given height
and frequency multiplied by the attenuation of the radiation (determined by $e^{-\tau_{\nu}/\mu}$) as it propagates outward in the direction of $\mu$.  The continuum optical depth $\tau_{\nu}$ is obtained by integrating the total continuum opacity ($\chi_{\nu}(z)$) over height.  The NLTE 
spontaneous bound-free (b-f) emissivity and b-f opacity corrected for stimulated emission \citep[Equations 7-1 and 7-2 of][]{Mihalas1978}
 are calculated using the NLTE populations computed in RADYN for a six-level hydrogen atom.  Other continuum transitions (involving higher levels of hydrogen, H$^-$, and metals\footnote{Although the population of H$^-$ is not considered in the equation of radiative transfer and level population equation, its population density is calculated from the NLTE densities of electrons and neutral hydrogen atoms.})
are calculated in LTE, as done internally in
RADYN.   We also consider the NLTE opacity and emissivity from 
induced recombination, Thomson scattering, and Rayleigh scattering.  

\begin{figure}[b] 
\begin{center}
 \includegraphics[width=5in]{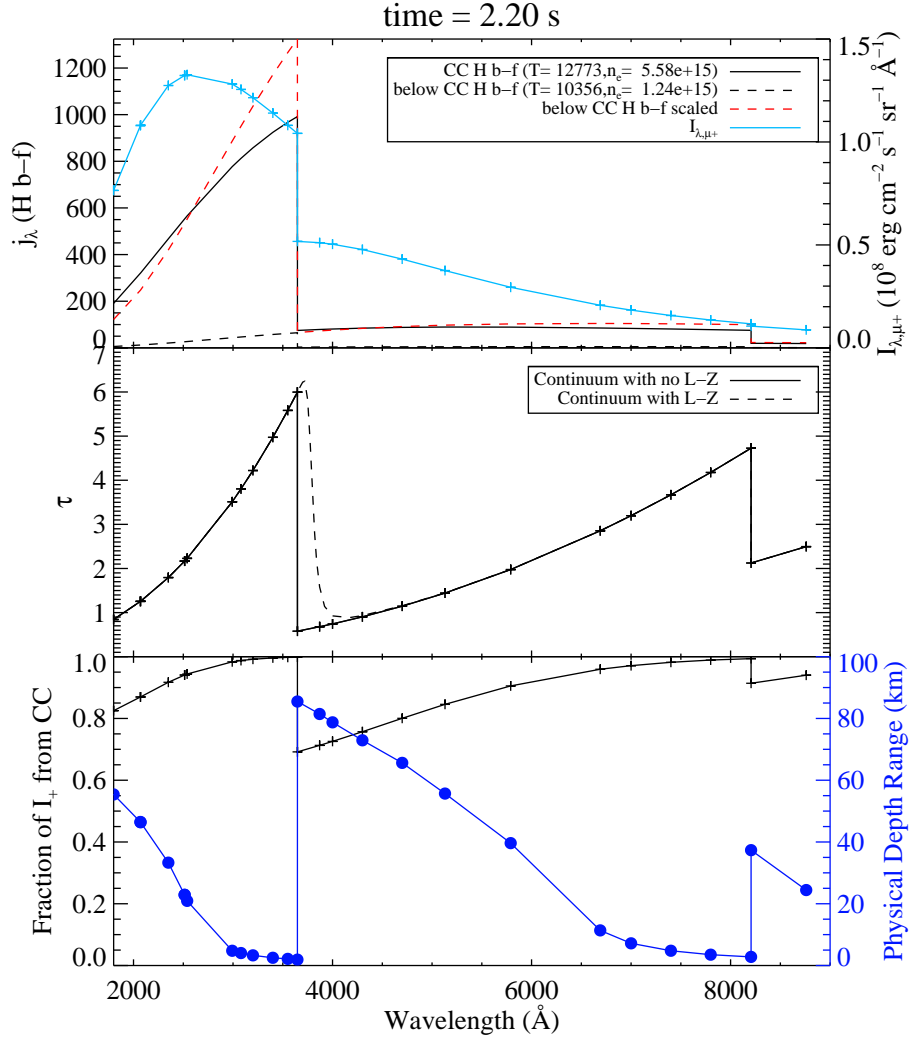} 
 \caption{The formation of emergent continuum intensity ($\mu=0.95$) in the F13 atmosphere at $t=2.2$~s.    (top):  The spontaneous hydrogen b-f emission
 at representative layers in the CC and in the stationary flare layers below.  The dashed red line shows the emissivity from the stationary layers that has been scaled 
 by a multiplicative factor so that the spectral variation is apparent.  The emergent intensity ($\mu=0.95$) is shown as the 
 light blue line (right axis).  (middle):  Optical depth ($\tau_{\nu}/\mu$) variation of the continuum, calculated at the bottom of the CC; the dashed curve 
 shows the continuum optical depth at wavelengths longer than the Balmer jump with Landau-Zener (L-Z) transitions included.  (bottom):  The fraction of emergent continuum intensity from the CC (black crosses, left axis) and the 
 physical depth range of continuum-emitting layers in the F13 atmosphere (blue circles, right axis).  The emergent intensity (light blue line in top panel) is approximately the appropriate emissivity in the top panel (with H f-f and H induced b-f emissivity spectra added) multiplied by the exponential
 of the negative value of the middle panel which is then integrated over the physical depth range in the bottom panel.   }
  \label{fig:panels}
\end{center}
\end{figure}

The emissivity and optical depth vary as a function of 
wavelength and depth in the atmosphere and thus lead to the properties of the 
spectral energy distribution of the emergent intensity.  The dominant emissivity in this model is
spontaneous hydrogen Balmer and Paschen recombination emissivity, with
lesser (but non-negligible) contributions from hydrogen free-free (f-f)
and induced hydrogen recombination.
The NLTE spontaneous hydrogen b-f emissivity spectra for representative layers in the CC
and the stationary flare layers are shown in the top panel of Figure \ref{fig:panels}.
In the middle panel, we show the total continuum optical depth ($\tau_{\nu}/\mu$) at the
bottom of the CC (where the downward-directed gas speed decreases below 5 km s$^{-1}$).   Hydrogen has large population
densities in the 
$n=2$ and $n=3$ levels in the CC,
which leads to optical depths of nearly $\tau\approx1$ at blue wavelengths and
optical depths $\tau>1$ at NUV and red wavelengths.
The optical depth variation reflects the hydrogenic b-f cross-section variation.

The emergent intensity spectrum (blue curve in top panel of Figure \ref{fig:panels}) can be understood as the integral of the
contribution function (Equation \ref{eq:didz}) over height where the emissivity at each height is
exponentially attenuated by the optical depth. 
 The hydrogen b-f emissivity spectrum decreases quickly in
the NUV, has a large Balmer jump ratio, and has a relatively flat spectrum between the blue (4000
\AA) and red (6690 \AA) wavelengths.  Between recombination edges, the wavelength dependence of the hydrogen b-f emissivity (in units of 
erg cm$^{-3}$ s$^{-1}$ sr$^{-1}$ \AA$^{-1}$)
is proportional to $\lambda^{-2} e^{\frac{-hc}{kT\lambda}}$, and 
the color temperature of the emissivity is related to the emissivity ratio at two wavelengths (through Equation 3 of
K15).  The color temperature
is $T\approx5000$ K for optically thin hydrogen recombination emission at
$T\approx10^4$ K, and $T\approx5000$ K would also correspond to the color temperature of the emergent
intensity if the hydrogen b-f emissivity (e.g., either emissivity curve in the top panel of Figure \ref{fig:panels}) originates only from layers with low optical depth.
Due to the large optical depths and the wavelength dependent variation of the optical depth in the CC (middle panel of Figure \ref{fig:panels}), the attenuation, given by $e^{-\tau_{\nu}/\mu}$,
multiplies by the top panel\footnote{The other non-negligible emissivities from H f-f and H induced b-f are first added to the H spontaneous b-f emissivity.} and results in the modified spectral energy distribution of the 
emergent intensity compared to the spectral energy distribution of the emissivity. 

Both the condensation and stationary flare layers contribute to the
spectral energy distribution of the 
emergent spectrum.  However, only the blue wavelengths (e.g., $\lambda=4300$ \AA) are (semi)transparent in the stationary flare layers because these wavelengths have the lowest optical depths.  Other wavelengths become opaque in the CC.  The fraction of the emergent intensity that originates
from the chromospheric condensation is shown in the bottom panel of
Figure \ref{fig:panels}): about 25\% of the blue continuum
radiation originates in the stationary layers below the CC.  Although the emissivity (top panel) is relatively low in these layers because of the lower 
electron density, the stationary flare layers have a larger vertical extent
and the integration of the contribution function over height gives a large value.    
We compare the transparency of photons using a physical depth range, $\Delta z(\lambda)$.
The (normalized) cumulative contribution function
$C_I^{\prime}$ is given as 

\begin{equation}
C_I^{\prime} (z,\mu) = \frac{\int_{z}^{z=10\rm{Mm}} C_I(z,\mu) dz}{I_\lambda(\mu)}
\end{equation}
where $ I_\lambda(\mu)$ is the emergent intensity and $z$ is the height variable (the height variable is defined as $z=0$ at $\tau_{5000}=1$; $z=10$ Mm corresponds to the top of the model corona).
Here we define the physical depth range\footnote{In K15, the physical depth range of
  the CC was calculated as the FHWM of the
  contribution function within the CC.} $\Delta z$ as the height difference between 
$C_I^{\prime}=0.95$ and $C_I^{\prime}=0.05$.  The value of $\Delta z$ thus defines the height
range over which the majority of the emergent intensity is formed.

In Figure \ref{fig:panels} (bottom), we show the physical depth range as a
function of wavelength.  For wavelengths in the 
NUV (3550\AA), blue (4300\AA), and red (6690\AA), the physical depth
ranges are $\Delta z = 2$, 72, and 11  km, respectively.  These are indicated qualitatively as vertical bars in
Figure \ref{fig:cartoon}, which illustrates that the physical depth range of
blue light extends to the stationary layers, whereas the physical
depth ranges of the opaque wavelengths are confined to the CC.
The larger transparency of blue photons leads to more photons that can escape
from the deeper layers, thus
increasing the blue emergent intensity relative to the NUV and red
intensities.  For a similar reason, the larger transparency of
$\lambda=3000$ \AA\ NUV photons compared to
$\lambda=3600$ \AA\ NUV photons produces the ratio of the emergent intensities
 (apparent from the light blue curve in the top panel of Figure
\ref{fig:panels}) that gives a hot color temperature at wavelengths $\lambda<3646$ \AA. At
$\lambda < 2500$ \AA, the spontaneous hydrogen b-f emissivity rapidly decreases (top
panel of Figure \ref{fig:panels}) and the larger physical depth range at these wavelengths is
not large enough to compensate, which results in a peak and turnover towards shorter wavelengths 
in the emergent intensity spectrum. 
The physical depth range variation of hydrogen
b-f (and f-f) emissivity thus produces an apparently hot color temperature of
$T=10,000$ K and a smaller Balmer jump ratio than are characteristic of optically
thin hydrogen recombination radiation.

The Balmer jump ratio and the color temperature of the continuum at
the blue and red wavelength sides of the Balmer jump are the result of large optical depths of the continuum
emission.  Surprisingly, the values of the color temperature of the continuum is not \emph{a priori} the temperature of the emitting plasma.  However,
the large optical depths directly result from atmospheric temperatures between $T=10,000 -
13,000$ K, which produces large thermal populations in 
the $n=2$ and $n=3$ levels of hydrogen.

\subsection{Opacities from Landau-Zener Transitions}
Using the RH code, we included the opacity effects from Landau-Zener transitions
of electrons within hydrogen atoms with upper energy
levels that overlap.  The modeling technique is employed in state-of-the art white dwarf
model atmosphere codes \citep{Tremblay2009}.  Perturbations from an electric microfield due to  ambient
charge density perturbs the highest energy levels of hydrogen, which is the
Stark effect.  At a critical microfield value \citep[which is determined by the
proton density;][]{HM88}, the highest Stark state of an upper level $n$ of hydrogen will 
overlap with the lowest Stark state of $n+1$, and all levels $\ge n$ are ``dissolved''. As a result,
bound-bound opacity is transferred to b-f opacity via a reverse cascade of electrons that undergo Landau-Zener
transitions among the dissolved upper levels.  Opacity effects are produced near the Balmer edge because the highest energy levels are
most easily perturbed and the energy separation converges for the upper principal quantum states.  
The modifications to the opacities resulting from Landau-Zener 
transitions give Balmer recombination emission at wavelengths longer than the
Balmer edge and results in a continuous transition from Balmer to Paschen
continuum opacity (and emissivity). 

 The Landau-Zener bound-free opacity at
$\lambda>3646$ \AA\ is the Balmer continuum opacity extrapolated longward of the
Balmer edge which is then multiplied by the dissolved fraction, $D(\lambda)$ \citep{Dappen1987}.
This opacity is added to the total continuum opacity longward of the Balmer
edge.  The resulting continuum optical depth at the bottom of the CC is shown as the dashed line in the middle panel
of Figure \ref{fig:panels}; the
wavelength extent of the extrapolation of the Balmer opacity longward
of the edge ($\lambda_{\rm{max -opacity}}-\lambda_{\rm{edge}}$)
is related to the ambient proton density in the flare
atmosphere (see Figure 9 of K15).  The maximum ambient proton density
attained in the F13 atmosphere is $n_p\approx5.6\times10^{15}$ cm$^{-3}$, and thus $\lambda_{\rm{max -opacity}} \approx 3730$ \AA.
The Landau-Zener transitions decrease the bound-bound opacity for transitions with dissolved upper levels,
which causes the flux of the highest order Balmer lines to fade into the
Balmer continuum emission that is produced redward of $\lambda=3646$ \AA.  Thus, the bluest Balmer line in emission
is also related to the ambient proton density.  In the F13 model at $t=2.2$~s, the
bluest identifiable Balmer line is H10 or H11, which is similar
to some impulsive phase spectra of dMe flares \citep{Kowalski2013, GarciaAlvarez2002}; in
other flares however, the H15 and H16 lines have been detected \citep{HawleyPettersen1991, Fuhrmeister2008, Kowalski2010}.


\section{Future Modeling Work}
 The F13 beam model reproduces the observed continuum flux ratios and the continuum
 properties within the complicated spectral region near the Balmer edge.
 These spectral properties can thus be used to infer the physical properties of
 the flare plasma:  continuum flux ratios (the Balmer jump ratio and the color temperature of the
continuum) can be used to infer the optical depth, and the wavelength extent of
continuum redward of the Balmer jump -- or the relative
intensity of a higher order Balmer line (H11-H16) -- can be used to infer the 
charge density of the continuum-emitting layers.

In \cite{Kowalski2013}, the instantaneous $t=2.2$~s F13 model was compared to 
time-resolved spectra during a large flare on the dM4.5e star YZ CMi; agreement
between the model and observations was found in the mid rise-phase
spectrum.  The burst-averaged
spectrum (over the 5~s simulation) gives a more direct comparison to
the observations, and was in better agreement with the 
early rise phase or the early gradual decay phase spectra.  In future
models, the persistence of bright continuum emission needs to be
extended for longer times using a sequence of flare bursts in order to be consistent with the observed
timescales of white-light flares, which can persist for
hours.  

The generation of CC's has long thought to be an important aspect of the standard solar
flare model, which includes the precipitation of electrons into the chromosphere producing
explosive mass motions \citep{Livshits1981, Fisher1985a, Fisher1985b, Fisher1989, Gan1992}.  The 
role of chromospheric condensations in solar and M dwarf flares is supported by observed redshifts
 in the chromospheric lines of H$\alpha$ \citep{Ichimoto1984, Canfield1990} and Mg \textsc{ii}
  \citep{Graham2015} and the transition region lines of Si \textsc{iv} and C \textsc{iv} \citep{Hawley2003}.  Thus, the origin of the 
white-light continuum in stellar flares in CC's (and the stationary layers just below)
 allows us to place this enigmatic phenomenon in the context of
the standard solar flare model, albeit with a larger energy flux
than usually considered for solar flares.  Beam fluxes ranging from $10^{11} - 10^{12}$ erg cm$^{-2}$ s$^{-1}$ 
were also considered in K15, but these fluxes (keeping
all else the same as the F13 beam parameters) produce an
unsatisfactory match to the observed continuum and emission line flux ratio properties.

We note that the F13 model predicts redshifted, broad chromospheric emission lines, such 
 as H$\alpha$.  Though there are only a few high-time resolution observations of 
M dwarf flares around H$\alpha$ with sufficient spectral resolution to infer mass motions, these typically show blueshifts of several tens of km s$^{-1}$ in the impulsive phase \citep{Hawley2003, Fuhrmeister2008}.  

A number of improvements will be made to our radiative-hydrodynamic flare modeling. First, the F13 beam
experiences significant energy loss from a return current electric
field, which is discussed in K15. In future models, we will incorporate the 
energy loss from the return current electric field into the updated RADYN flare code \citep{Allred2015} and compare to the continuum emission 
produced by the F13 beam in K15.  We will also consider an M dwarf flare model with a
 higher low-energy cutoff in the electron energy distribution, which mitigates return current effects.
We plan to implement an improved prescription for Stark broadening
using the \cite{Vidal1973} theory in place of the analytic approximations that are currently
used (see K15 for discussion).  Thus, we will rigorously determine if charge densities
attained in M dwarf flares are as high as $n_e \approx 5\times10^{15}$ cm$^{-3}$.  Future observations with high cadence echelle flare spectra around the Balmer jump
and H$\alpha$ can provide critical constraints for the formation of such dense CC's in M dwarf flares.  If high density CC's are formed in M dwarf flares, 
then we may need to consider a physical mechanism, such neutral beams \citep{Karlicky2000}, re-acceleration processes \citep{Varady2014}, or in-situ chromospheric acceleration \citep{Fletcher2008}, to allow F13 beams to penetrate into the lower atmosphere.

 \section{Implications for Superflares}
 Recently, superflares in rapidly rotating G stars have been observed with Kepler, revealing 
 energies in the white-light continuum of $E\approx10^{35}-10^{36}$ erg \citep{Maehara2012, Shibayama2013, Maehara2015}.  
 These energies are comparable to or even larger than the largest flares observed in M dwarf stars \citep{HawleyPettersen1991, Osten2010, Kowalski2010, Schmidt2014, Drake2014}.
 Thus, the energy and duration of the white-light superflares in G stars pose a similar challenge 
 for flare models.  Recently, lower nonthermal electron beam energy fluxes of $3\times10^{11}$ erg cm$^{-2}$ s$^{-1}$ have been used to model 
 the areal extent of the white-light emission in superflares \citep{Katsova2015}, but NUV/optical spectra during a superflare
 would be invaluable for determining if higher energy fluxes are necessary.  These spectra would 
 provide new constraints on the color temperature, the Balmer jump ratio, and the properties of the highest order Balmer lines to be directly compared to flare spectra of M dwarf flares.
 Despite the large peak luminosities, superflares in G stars produce a low contrast against the background emission.  Therefore, obtaining high signal-to-noise flare spectra (such as those 
 obtained during a flare on the early type K dwarf AB Doradus from \cite{Lalitha2013}) will be an important observational challenge in the future. 
 
 \section*{Acknowledgments}
 AFK thanks the organizers of IAUS \#320 for the opportunity to present this work.   AFK thanks S. Hawley, K. Shibata, L. Fletcher, G. Cauzzi, and P. Heinzel for stimulating discussions at IAUS \#320.
AFK acknowledges travel support from HST GO 13323 from the Space Telescope Science Institute, operated by the Association of Universities for Research in Astronomy, Inc., under NASA contract NAS 5-26555, and for travel support from the International Astronomical Union.

\clearpage
\bibliography{adam_iaus320}

\end{document}